# Equivalent Circuit Representation and Thermodynamic Limitations of Non-Adiabatic Spin-Transfer Torque effect

Wataru Koshibae

*RIKEN Center for Emergent Matter Science (CEMS), Wako, 351-0198, Japan*

We present an equivalent circuit representation of the Thiele equation for the current-driven domain walls. We show that the non-adiabatic spin-transfer torque (β-term) cannot be represented by passive elements alone (such as resistors, inductors, and capacitors) and necessarily requires active elements, corresponding to negative dissipation. Consequently, assuming that the driving current is the only power source for the texture dynamics, directly incorporating the β-term introduces a fundamental contradiction with thermodynamic principles. This equivalent circuit theory indicates that the interpretation of experiments solely through the β-term does not fully capture the underlying physics, highlighting the need to account for effects distinct from the spin dynamics.

## I. Introduction

Current-driven dynamics of magnetic domain walls have attracted extensive attention due to their fundamental physical interest and potential applications [1-5]. To date, the momentum transfer from conduction electrons to localized spins has been categorized into two mechanisms: the adiabatic spin-transfer torque (STT) and the non-adiabatic STT, commonly characterized by the dimensionless parameter β [4-12]. To describe the collective motion of the localized magnetic configurations, the Thiele equation [13]—derived from the Landau-Lifshitz-Gilbert (LLG) equation—has long served as a standard analytical framework. While the β-term plays a key role in determining dynamical features, such as the critical current threshold and the steady-state drift velocity, its microscopic origin and theoretical consistency within phenomenological models remain a subject of active debate.

A promising approach to clarifying such phenomenological models is equivalent circuit analysis, which maps complex dynamical systems onto electrical circuits—a technique also employed beyond physics [14-16] to model system dynamics using active elements. The application of the equivalent-circuit method to the LLG equation is particularly useful for systems involving nonconserved currents, such as spin currents. The equivalent-circuit representation not only provides intuitive insights into system behavior but also establishes a rigorous connection between magnetization dynamics and circuit theory. In particular, circuit theory imposes strict thermodynamic constraints on physical systems, such as the requirement of passivity for systems composed of passive elements. When applied to magnetic textures, an equivalent circuit representation of the Thiele equation provides a clear framework for examining whether current-driven torques satisfy these thermodynamic considerations.

In this paper, we construct an equivalent circuit representation of the Thiele equation for the domain wall dynamics to investigate the thermodynamic validity of current-driven spin torques. Through equivalent circuit analysis, we demonstrate that the non-adiabatic STT (β-term) cannot be modeled using passive circuit elements alone and inherently requires active elements corresponding to negative dissipation. Consequently, assuming that the driving current is the only power source for the texture dynamics, directly incorporating the β-term introduces a fundamental contradiction with thermodynamic principles. This paper is organized as follows. In Section II, we derive the equivalent circuit for the Thiele equation, formulate its impedance, and analyze the passivity/activity characteristics along with the thermodynamic contradiction arising from the β-term. Finally, Section III concludes the paper.

**II. Formulation of Impedance via Spin Electromotive Force**

The LLG equation for the current-driven spin dynamics in the standard notations is expressed by

$$\frac{\partial \boldsymbol{n}}{\partial t} = -\gamma \boldsymbol{n} \times \boldsymbol{h} + \alpha \boldsymbol{n} \times \frac{\partial \boldsymbol{n}}{\partial t} - (\boldsymbol{v_s} \cdot \boldsymbol{\nabla})\boldsymbol{n} + \beta \boldsymbol{n} \times [(\boldsymbol{v_s} \cdot \boldsymbol{\nabla})\boldsymbol{n}] \tag{1}$$

where $\boldsymbol{n}$ is the unit vector along the local magnetization, $\gamma\ (> 0)$ is the gyromagnetic ratio, $\boldsymbol{h}$ characterizes the underlying magnetic interactions of the system, $\alpha$ is the Gilbert damping constant, $\beta$ is the non-adiabatic parameter, and $\boldsymbol{v_s}$ is the spin-drift velocity proportional to the applied current density.

For a one-dimensional magnetic domain wall, integration of the LLG equation Eq.(1) yields the equation of motion:

$$\begin{cases} \frac{1}{\Delta}\dot{q} - v_s\frac{1}{\Delta} - \alpha\dot{\phi} = +\frac{1}{2\Delta}\gamma\frac{1}{M_s A}\frac{\partial U_{ani}}{\partial \phi} \\ \dot{\phi} - \beta v_s\frac{1}{\Delta} + \alpha\frac{1}{\Delta}\dot{q} = -\frac{1}{2}\gamma\frac{1}{M_s A}\frac{\partial U_{pin}}{\partial q} \end{cases} \tag{2}$$

where $q$ and $\phi$ represent the domain wall position and its internal magnetization angle (azimuthal tilt angle), respectively, with the domain wall width $\Delta$, the magnitude of the local magnetization $M_s$, and the cross-sectional area of the sample $A$. The anisotropy and impurity potentials are denoted by $U_{ani}$ and $U_{pin}$, respectively.

Below the Walker breakdown threshold current, we apply the harmonic approximation to the anisotropy potential,

$$U_{ani} = \frac{1}{2} k_\phi \phi^2. \tag{3}$$

For the extrinsic pinning effect, we also assume

$$U_{pin} = \frac{1}{2} k_q q^2 \tag{4}$$

in this paper.

Applying the Fourier transform, we find the solutions of Eq.(2),

$$\Phi_\omega = \frac{\left(-i\omega\alpha - \Delta \frac{\gamma}{2M_s A} k_q\right)\left(-\mathrm{v}_{s,\omega}\right) + i\omega\left(-\beta \mathrm{v}_{s,\omega}\right)}{\left(i\omega\Delta\alpha + \frac{\gamma}{2M_s A} k_\phi\right)\left(-i\omega\alpha - \Delta \frac{\gamma}{2M_s A} k_q\right) + \omega^2 \Delta} \tag{5}$$

and

$$\mathrm{q}_\omega = \frac{i\omega\Delta\left(-\mathrm{v}_{s,\omega}\right) + \left(i\omega\Delta\alpha + \frac{\gamma}{2M_s A} k_\phi\right)\left(-\beta \mathrm{v}_{s,\omega}\right)}{\left(i\omega\Delta\alpha + \frac{\gamma}{2M_s A} k_\phi\right)\left(-i\omega\alpha - \Delta \frac{\gamma}{2M_s A} k_q\right) + \omega^2 \Delta} \tag{6}$$

where $\Phi_\omega, \mathrm{q}_\omega$ and $\mathrm{v}_{s,\omega}$ are the Fourier transform of $\phi, q$ and $v_s$, respectively.

The spin-drift velocity $v_s$ is related to the applied electric current $I$ with a magnetic polarization $P$ and standard notations,

$$v_s = -\left(\frac{g\mu_B P}{2|e| M_s A}\right) I = -\left(\hbar \frac{P}{|e|} \frac{\gamma}{2M_s A}\right) I \equiv -|b| I \tag{7}$$

and therefore, $\mathrm{v}_{s,\omega} = -|b| \mathrm{I}_\omega$ where $\mathrm{I}_\omega$ is the Fourier transform of $I$. The spin electromotive force $V$ [17] is expressed as,

$$V = P \frac{\hbar}{|e|} \dot{\phi} = \frac{2M_s A}{\gamma} |b| \dot{\phi} \tag{8}$$

and therefore, the Fourier transform of $V$ is given by,

$$\mathrm{V}_\omega = \frac{2M_s A}{\gamma} |b| i\omega \Phi_\omega . \tag{9}$$

As a result, the solution of $\Phi_\omega$ (Eq.(5)) is expressed as $\mathrm{V}_\omega = \mathrm{V}_{\text{adiab.}} + \mathrm{V}_{\text{non-adiab.}}$ with

$$\mathrm{V}_{\text{adiab.}} = \frac{1}{\frac{1}{R_{1/\alpha}} + \frac{1}{i\omega L} + \frac{1}{R_\alpha + \frac{1}{i\omega C}}} \mathrm{I}_\omega \tag{10}$$

and

$$\mathrm{V}_{\text{non-adiab.}} = \frac{1}{-\frac{1}{R_\beta}\left(\frac{1}{R_{1/\alpha}} + \frac{1}{i\omega L}\right)\left(R_\alpha + \frac{1}{i\omega C}\right) - \frac{1}{R_\beta}} \mathrm{I}_\omega \tag{11}$$

where $R_{1/\alpha} = \frac{1}{\alpha} \frac{2M_s A}{\gamma} \frac{|b|^2}{\Delta}$, $R_\alpha = \alpha \frac{2M_s A}{\gamma} \frac{|b|^2}{\Delta}$, $L = \left(\frac{2M_s A}{\gamma}\right)^2 \frac{|b|^2}{k_\phi}$, $C = \frac{1}{|b|^2 k_q}$, and $R_\beta =$

$\beta\frac{2M_s A}{\gamma}\frac{|b|^2}{\Delta}$ . The sharp contrast of $V_{\mathrm{non-adiab.}}$ to $V_{\mathrm{adiab.}}$ is characterized by $R_\beta$, i.e., in the case that $\beta = 0$, $V_{\mathrm{non-adiab.}} = 0$. This result reveals that the effect of the non-adiabatic STT effect (β-term) is entirely captured within $V_{\mathrm{non-adiab.}}$

Figure 1 shows the equivalent circuit representation of $V_{\mathrm{adiab.}}$. As seen in Fig.1, the adiabatic STT responses are expressed by the passive elements. The roles of α, $k_\phi$ and $k_q$ are transparent: The Gilbert damping constant α gives the resistances $R_{1/\alpha}$ and $R_\alpha$. The coefficient $k_\phi$ represents the intrinsic pinning effect and this results in the inductance $L$. On the other hand, the extrinsic pinning effect $k_q$ is described by the capacitance $C$.

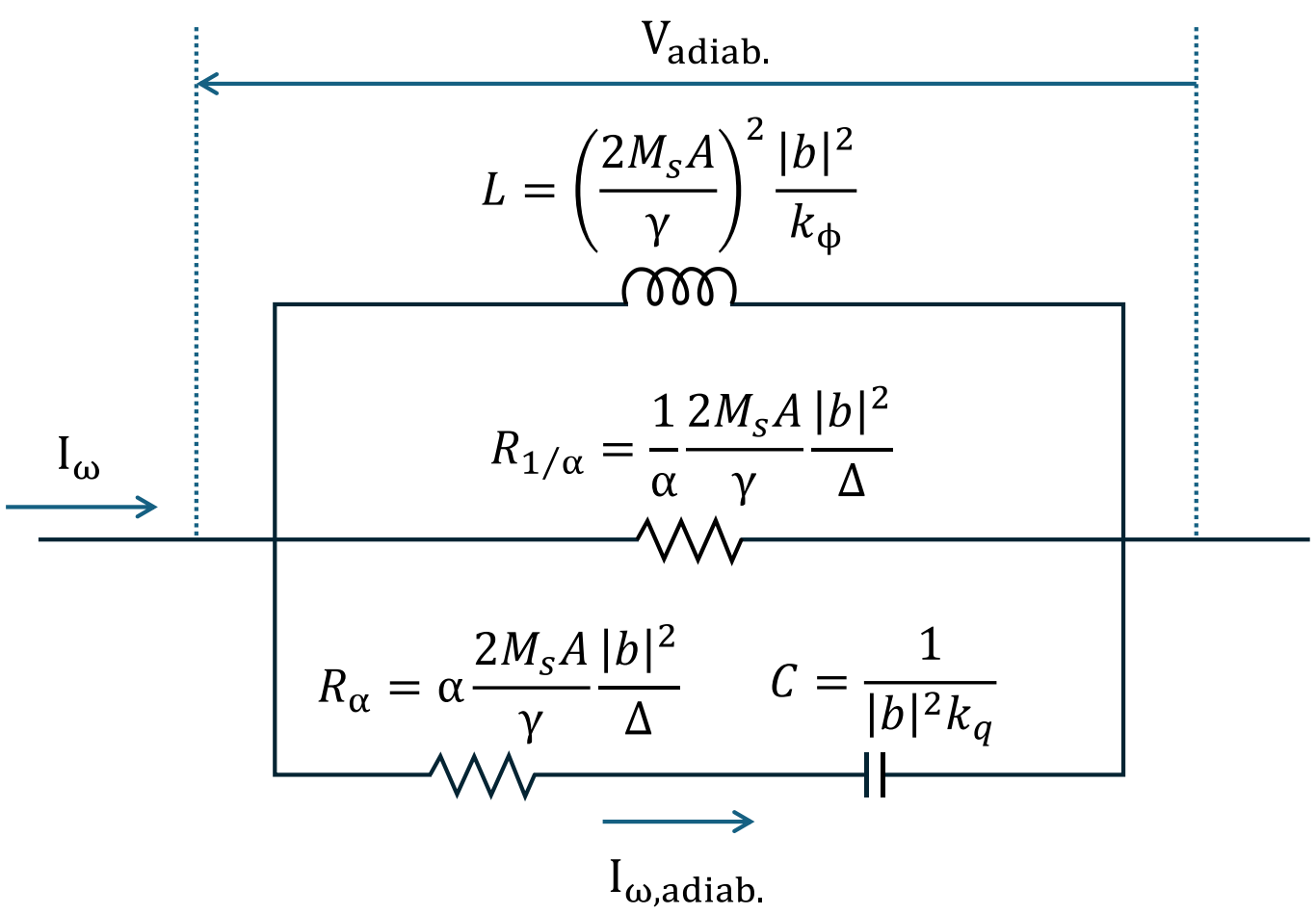

Fig. 1 The equivalent circuit representation of $V_{\mathrm{adiab.}}$. The arrows define the positive-sign directions of the voltage and the currents. The interpretation of the current on the bottom branch is discussed later.

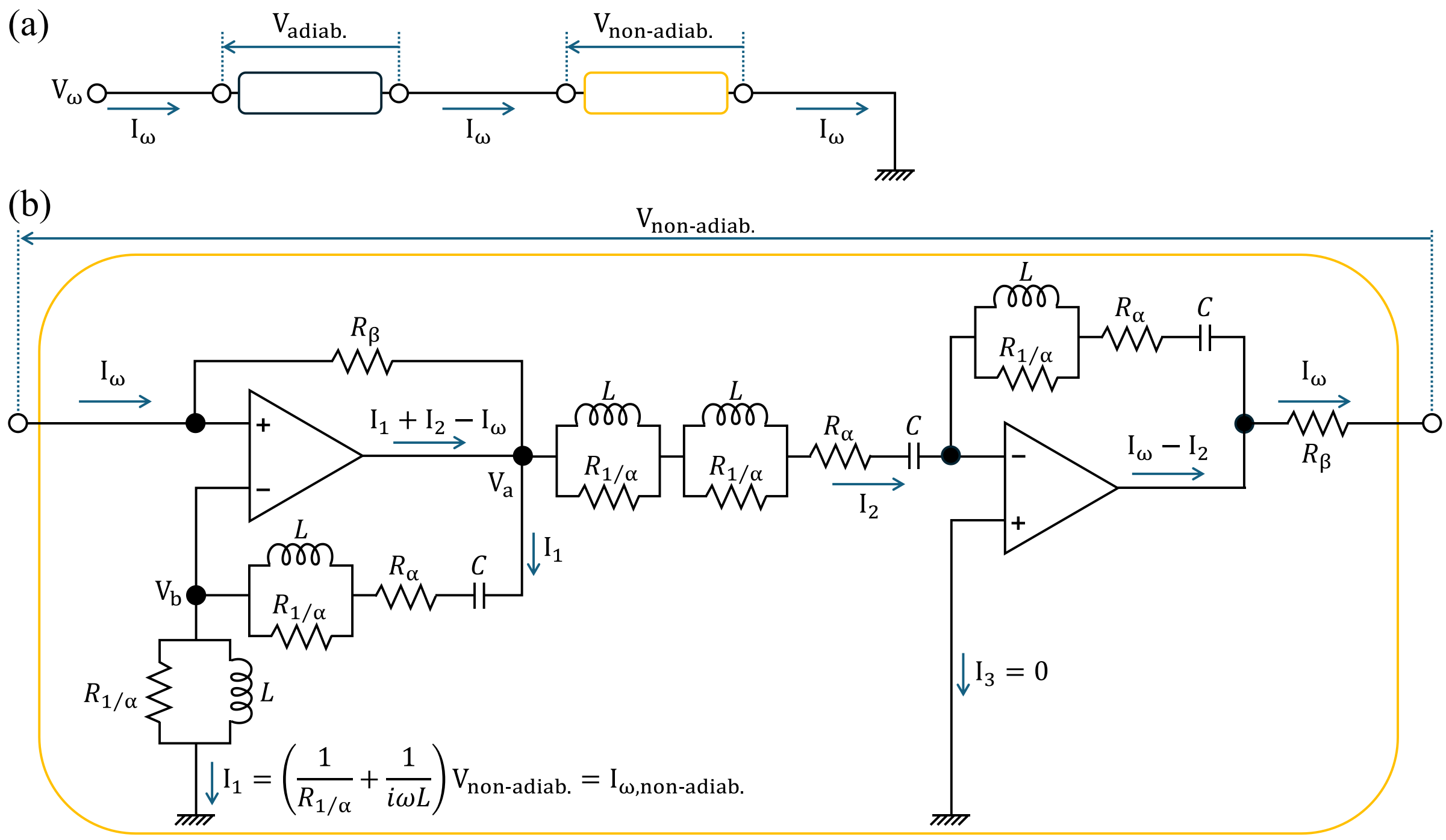

Fig. 2 (a) A model circuit. (b) An equivalent circuit representation of $V_{non-adiab.}$. The arrows define the positive-sign directions of the voltages and the currents. The interpretation of the current $I_{\omega,non-adiab.}$ is discussed later.

The expression of $V_{non-adiab.}$ (Eq.(11)) clearly indicates that such a response is impossible to synthesize with conventional passive elements, thereby requiring active components. Figure 2 shows an equivalent circuit representation of $V_{non-adiab.}$. This circuit combines a basic Negative Impedance Converter (NIC) structure with an inverting amplifier, both of which are standard operational amplifier (op-amp) building blocks [18]. Note that, as seen in Fig.2(b), input current at the left-end terminal is $I_\omega$, but output currents are $I_\omega$ at the right-end terminal and $I_1$ indicated below the left op-amp. This current $I_1$ flows along a path entirely distinct from that of the main driving current $I_\omega$. The current source of $I_1$ is the op-amps, i.e., $(I_1 + I_2 - I_\omega) + (I_\omega - I_2) = I_1$. Specifically, under the virtual short condition of the left op-amp in Fig.2(b), i.e., $V_{non-adiab.} = V_b$, Fig.3 illustrates the current-voltage (I-V) characteristics for $I_1$:

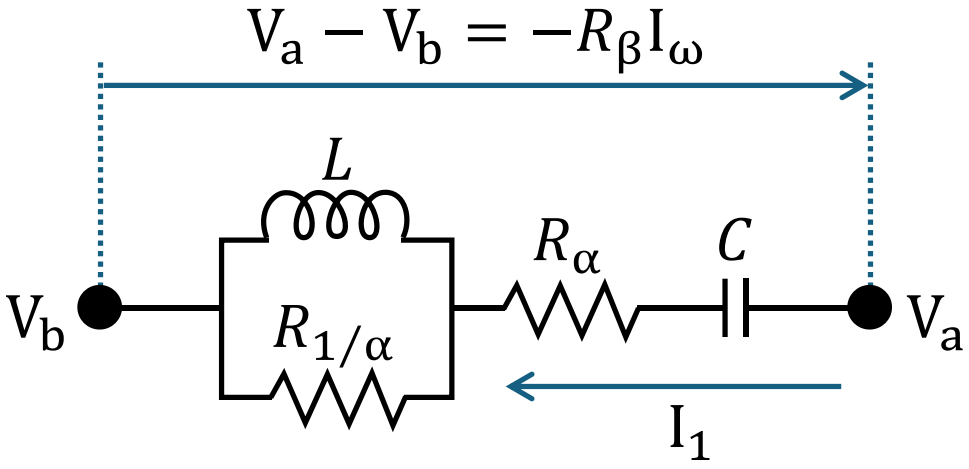

Fig. 3 The current-voltage (I-V) characteristics for $I_1$ in the circuit shown in Fig.2(b). The arrows define the positive-sign directions of the voltage and the current.

Therefore, the origin of $I_1$ is the power source to operate the op-amps entirely distinct from that of the main driving current $I_\omega$. Figure 3 clearly shows that $R_\beta$ controls $I_1$, i.e., $I_1 \to 0$ in the case that $\beta \to 0$. In this limit, the left and right terminals of the circuit in Fig.2(b) *behave* as if they were short-circuited ($V_{\text{non-adiab.}} \to 0$). Note that this short-circuit *behavior* in the limit $\beta \to 0$ requires the op-amp to be powered, whereas the circuit for $V_{\text{non-adiab.}} = 0$ at $\beta = 0$ is naturally given by a true physical short circuit (see Eqs. (5) and (11)).

Next, let us discuss the solution $q_\omega$ (Eq.(6)). The Fourier transform of the domain wall velocity $v_q = \dot{q}$ is expressed as $v_{q,\omega} = i\omega q_\omega$. Therefore, Eq.(6) indicates that the domain wall velocity is given by the algebraic sum of the contributions from the adiabatic and non-adiabatic torques, i.e.,

$$v_{q,\omega} = -|b| I_{\omega,\text{adiab.}} + |b| I_{\omega,\text{non-adiab.}} \tag{12}$$

with

$$I_{\omega,\text{adiab.}} = \frac{1}{R_\alpha + \frac{1}{i\omega C}} V_{\text{adiab.}} \tag{13}$$

and

$$I_{\omega,\text{non-adiab.}} = \left(\frac{1}{R_{1/\alpha}} + \frac{1}{i\omega L}\right) V_{\text{non-adiab.}}\,. \tag{14}$$

Interestingly, $I_{\omega,\text{adiab.}}$ is corresponding to the current of the bottom branch of the circuit shown in Fig.1. At the same time, on the path entirely distinct from that of $I_{\omega,\text{adiab.}}$, $I_{\omega,\text{non-adiab.}} = I_1$ in Fig.2. In other words, the current $I_{\omega,\text{non-adiab.}}$ is *not* driven by the main driving current $I_\omega$, but comes from the independent power supply of the op-amp circuit.

In DC limit, $\omega \to 0$, the circuit shown in Fig.1 immediately concludes that $I_{\omega\to 0,\text{adiab.}} = 0$. Thus, $v_{q,\omega\to 0} = +|b| I_{\omega\to 0,\text{non-adiab.}} = +|b| I_1$. On the other hand, Fig. 3 provides the following solutions in DC limit, $\omega \to 0$:

(Case 1) $1/i\omega C = 0$, i.e., the case that $k_q = 0$. In this case, $I_{\omega\to 0,\text{non-adiab.}} = I_1 = (V_a - V_b)/R_\alpha = -(R_\beta/R_\alpha) I_{\omega\to 0}$. Therefore, Eq.(12) gives the solution $v_q = (\beta/\alpha) v_s$ which is consistent with Eq.(2).

(Case 2) $1/i\omega C \neq 0$, i.e., finite $k_q$ case. In Fig.3, the current $I_1(t)$ being the inverse Fourier transform of $I_1 = I_{\omega,\text{non-adiab.}}$ is expressed to be $I_1(t) = \dot{Q}_1$ where $Q_1$ is the charge stored in the capacitor $C$ in Fig.3. Because $v_{q,\omega} = i\omega q_\omega = +|b| I_1$, the charge $Q_1$ is related to the DW position $q$ by $q = +|b| Q_1$. In DC limit, $\omega \to 0$, $Q_1 = C(V_a - V_b) = -CR_\beta I$. Using Eq.(7), we find $q = CR_\beta v_s = \left(\frac{1}{|b|^2 k_q}\right)\left(\beta \frac{2M_s A}{\gamma} \frac{|b|^2}{\Delta}\right) v_s = \beta \frac{2M_s A}{\gamma k_q \Delta} v_s$. This is the circuit understanding of the solution of Eq.(2). Note that the charge $Q_1$ discussed here originates from the independent power supply for the op-amp circuit distinct from that of the driving current $I$.

As discussed above, by the equivalent circuit considerations, the spin dynamics driven by β originates solely from an independent power supply, which is entirely separate from the driving current $I_\omega$, i.e., the applied spin drift velocity in the adiabatic torque term ($-(\boldsymbol{v_s} \cdot \boldsymbol{\nabla})\boldsymbol{n}$) of the LLG equation Eq.(1). The op-amps cannot operate without such independent power supply distinct from the driving spin drift velocity $v_s$. The self-sustained active op-amp circuits correspond to the active behavior driven by the β-term, which contradicts thermodynamic principles. As also shown by the above discussion, the need to power the op-amp in the limit $\beta \rightarrow 0$ indicates that an infinitesimal β violates thermodynamic principles.

**III. Summary and Discussion**

This study clarifies the physical nature of the current-driven domain wall dynamics using an equivalent circuit representation. The equivalent circuit clearly illustrates the respective roles of the resistor (*R*), inductor (*L*), and capacitor (*C*) [19-22]. We have demonstrated that the non-adiabatic STT (β-term) cannot be synthesized purely by passive circuit elements (*R*, *L*, and *C*), but inherently require active elements. The circuit block for the β-term includes an NIC circuit, which corresponds to negative dissipation. By explicitly modeling an equivalent circuit using op-amps, we revealed that the actual energy source driving the domain wall dynamics via the β-term is not the primary driving current ($v_s$) of the LLG equation Eq.(1), but rather the external power supply required to operate the active devices. Such external power, via active elements, induces negative resistance (negative dissipation) as well as negative inductance and/or negative capacitance. At the same time, the β-term drives the domain wall motion, but not through the driving current $v_s$ itself. This finding indicates that assuming the driving current $v_s$ to be the sole energy source while directly incorporating the β-term leads to a fundamental contradiction with thermodynamic principles. Consequently, interpretation of the experiments solely through the β-term lacks a clear physical basis, highlighting the essential need to account for energy sources and/or physics distinct from the intrinsic spin dynamics.

As shown in Fig.1, the equivalent circuit of $V_{\mathrm{adiab.}}$ is expressed by the passive elements *R*, *L*, and *C*. Therefore, for the circuit, we can apply the thermodynamics understanding within the framework of linear response theory [23,24]. The dissipation occurs at $R_{1/\alpha}$ and $R_\alpha$. On the other hand, in the LLG equation Eq. (1), the energy dissipation rate $\partial_t E_{diss}$ due to α for a one-dimensional magnetic domain wall is represented by a Rayleigh dissipation function,

$$\frac{\partial E_{diss}}{\partial t} = -\alpha\frac{M_s A}{\gamma}\int\left|\frac{\partial \boldsymbol{n}}{\partial t}\right|^2 dx = -\frac{\alpha}{1+\alpha^2}\frac{M_s A}{\gamma}\int|\gamma\boldsymbol{n}\times\boldsymbol{h} + v_s\partial_x\boldsymbol{n}|^2 dx\,. \tag{15}$$

The consistency of our equivalent circuit model is easily verified. For example, let us consider the case under an AC $v_s$ with the high-frequency limit, $\omega \to \infty$. In this limit, $\boldsymbol{n}\times\boldsymbol{h} \to \boldsymbol{0}$, because the magnetic moment oscillation is unable to follow the high-ω. As a result, for the domain wall, $\partial_t E_{diss} = -\frac{\alpha}{1+\alpha^2}\frac{2M_s A}{\gamma\Delta}{v_s}^2$ which is fully understood through the circuit shown in Fig.1, i.e., the parallel circuit configuration of $R_{1/\alpha}$ and $R_\alpha$ with $|i\omega L| \to \infty$ and $|1/i\omega C| \to 0$ in the limit $\omega \to \infty$. In contrast, the equivalent circuit for the β-term shown in Fig.2 incorporates active elements, meaning that such linear response theory considerations cannot be applied. Our equivalent circuit approach provides a clear interpretation of the negative dissipation using the active elements, i.e., the NIC structure by op-amp in the equivalent circuit, which generates negative resistance (as well as negative inductance and/or negative capacitance). Thus, our findings reveal that the β-driven dynamics lies beyond the Rayleigh-Lagrangian framework. Because a Rayleigh dissipation function is formulated on the passivity of linear response theory—consistent with the fluctuation-dissipation theorem—it cannot be applied to systems driven by active elements. In addition to this, our theory also clarifies that the β-term driven domain wall dynamics originates from the independent energy source distinct from the driving current $v_s$. Such independent energy source, however, is not accounted for in the LLG equation Eq. (1).

From a mathematical and computational perspective, however, the β-term serves as a highly useful tool. As also discussed in the previous studies [8,9], in the absence of extrinsic pinning effects, setting $\alpha = \beta$ restores a remarkable symmetry—Galilean invariance—in the highly nonlinear LLG equation. This special condition provides an invaluable benchmark for both analytical treatments and numerical simulations, offering a rigorous criterion to verify the validity of derived solutions. By establishing a solid baseline under the mathematically well-behaved condition $\alpha = \beta$, one can systematically extend the analysis to more general β-dependent behaviors. In fact, numerical simulations of the LLG equation Eq.(1) confirm the validity of the equation of motion Eq.(2).

Nevertheless, this mathematical utility does not justify using the β-term to interpret experimental results. Our equivalent circuit analysis demonstrates that attributing measured domain wall dynamics solely to the β-term ignores the necessity of active energy sources, violating thermodynamic principles. Therefore, experimental data must not be interpreted through the β-term; instead, other distinct physical mechanisms—such as extrinsic thermal effects, or alternative torque contributions—must be carefully examined.

**Acknowledgment**

The authors are grateful to N. Nagaosa for fruitful and stimulating discussions. The authors acknowledge the use of generative AI tools (ChatGPT, Gemini, Rakuten AI, Claude, and Microsoft Copilot) for language proofreading and preliminary calculation checks. This work was supported by JSPS KAKENHI Grant No. 23K03291.

## Appendix

By employing an ansatz for a current-driven helical texture with wavelength $\lambda$, where the phase is given by $(2\pi/\lambda)(x-q)$ and $\phi$ represents the tilt angle with respect to the spin rotation plane at the ground state, the resulting equations of motion are isomorphic to those for current-driven domain walls. Therefore, the conclusion remains the same as in the main text.